\documentclass[journal, letter, 9pt]{IEEEtran}
\usepackage{amsmath}
\usepackage{amsfonts}
\usepackage{graphics}
\usepackage{array}
\usepackage{shortvrb}
\usepackage{epsf}
\usepackage{graphicx}
\usepackage{rotating}
\usepackage{multirow}
\usepackage{cite}
\usepackage{xcolor}
\usepackage{alltt}
\usepackage{soul}
\usepackage{overpic}
\usepackage{pict2e}
\usepackage{comment}
\usepackage{acronym}
\usepackage{float}
\usepackage[caption=false,font=footnotesize]{subfig}
\usepackage{placeins}
\usepackage{threeparttable}
\usepackage{booktabs}
\newcommand{\bfg}[1]{\boldsymbol{#1}}
\newcommand{\bfb}[1]{\boldsymbol{\rm #1}}

\acrodef{cig}[CIG]{converter-interfaced generator}
\acrodef{coi}[COI]{Center of Inertia}
\acrodef{avr}[AVR]{Automatic Voltage Regulator}
\acrodef{agc}[AGC]{Automatic Generation Control}
\acrodef{lic}[LIC]{Local Integration Control}
\acrodef{pfc}[PFC]{Primary Frequency Control}
\acrodef{sm}[SM]{synchronous machine}
\acrodef{der}[DER]{Distributed Energy Resource}
\acrodef{ffr}[FFR]{Fast Frequency Regulation}
\acrodef{pll}[PLL]{Phased-Locked Loop}
\acrodef{rocof}[RoCoF]{Rate of Change of Frequency}
\acrodef{dae}[DAE]{Differential-Algebraic Equation}
\acrodef{pi}[PI]{Proportional-Integral}
\acrodef{pod}[POD]{Power Oscillation Damper}
\acrodef{pf}[PF]{Participation Factor}
\acrodef{lep}[LEP]{Linear Eigenvalue Problem}
\acrodef{ess}[ESS]{Energy Storage system}
\begin{document}
\title{Dynamic Slack Bus}
\newcommand{\T}{^{\scriptscriptstyle \rm T}}
\newcommand{\wcoi}{\omega_{\scriptscriptstyle \rm COI}}
\newcommand{\dwcoi}{\dot\omega_{\scriptscriptstyle \rm COI}}
\newcommand{\vd}{v_{d}}
\newcommand{\vq}{v_{q}}
\newcommand{\dvd}{\dot{v}_{d}}
\newcommand{\dvq}{\dot{v}_{q}}
\newcommand{\vdt}{v'_{d}}
\newcommand{\vqt}{v'_{q}}
\newcommand{\dvdt}{\dot{v}'_{d}}
\newcommand{\dvqt}{\dot{v}'_{q}}
\newcommand{\wt}{\tilde{\omega}}

\newcommand{\ra}[1]{\renewcommand{\arraystretch}{#1}}

\newcommand{\Dt}{\frac{\rm d}{{\rm d}t}}
\newcommand{\ds}[1]{#1_{s, {\rm d}}}
\newcommand{\qs}[1]{#1_{s, {\rm q}}}
\newcommand{\dr}[1]{#1_{r, {\rm d}}}
\newcommand{\qr}[1]{#1_{r, {\rm q}}}
\newcommand{\dx}[1]{#1_{\rm d}}
\newcommand{\qx}[1]{#1_{\rm q}}
\newcommand{\wbi}{\omega_b^{-1}}
\newcommand{\dqi}[2]{\bar{#1}_{#2, {\rm dq}}}
\newcommand{\DQ}{{\rm dq}}
\newcommand{\dqti}[2]{\bar{#1}_{#2, {\rm dq}}(t)}
\newcommand{\Dpa}{\frac{\rm d}{{\rm d}t} + j\omega_o}
\newcommand{\0}{o}

\newcommand{\jj}{\jmath}

\author{ %
  Federico Milano, {\em IEEE Fellow} %
  \thanks{F.~Milano is with the School of Electrical and Electronic
    Engineering, University College Dublin, Belfield, Dublin, D04V1W8,
    Ireland. e-mail: \mbox{federico.milano@ucd.ie}%
  } %
  \vspace{-5.75mm}
}

\markboth{Submitted to IEEE PES Letters}{}
\maketitle
\begin{abstract}
  This letter proposes a general dynamic formulation of slack bus.  With this aim, the angle constraint imposed by the slack bus is redefined as a set of differential equations and an energy source.  The existence and role of the transient component of this source is also discussed in the letter.  Based on this framework, the letter shows that the swing equations of synchronous machines can be interpreted as distributed, dynamic, multi-variable, local slack buses.  Other relevant cases, including primary and secondary frequency regulation, passive loads as well as grid following and grid forming converters are discussed.
\end{abstract}
\IEEEpeerreviewmaketitle

\section{Introduction}
\label{sec:intro}

The concept of slack bus, its modeling and link with secondary frequency regulation or market dispatch is a recurring topic in power system analysis.  An in-depth review of the history and implementations of the slack bus model can be found in the introduction of \cite{Dhople20}.  In conventional power flow analysis, the slack bus model serves two purposes: (i) fix a reference phase angle; and (ii) provide a slack power to balance generation production, load consumption and network losses:
\begin{align}
  \label{eq:1}
  0 &= \theta^o - \theta_i \, , \\
  \label{eq:2}
  p_{h} &= p^o_{h} + k_{g, h} \, \hat{\sigma} \, , \qquad h = 1, \dots, n \, ,
\end{align}
where $i$ is the index of the $i$-th bus of the network; $p_{h}$ is the active power injection at the $h$-th bus; $k_{g, h}$ is a factor that indicates the participation of the $h$-th generator to the system losses; and $\hat{\sigma} \in \mathbb{R}$ is the ``slack'' variable.   The single bus formulation is obtained from \eqref{eq:2} by imposing $k_{g, h} = 0$ for $n-1$ generators.

The objective of this letter is to propose a general framework  to define and classify all possible static and dynamic slack bus models.  The letter also shows that the conventional slack bus model for power flow analysis is a special case of a more general dynamic formulation.  The motivation for the proposed general framework is twofold: (i) provide a precise mathematical definition of slack bus; and (ii) facilitate a better understanding of the functioning of power system devices, in particular, those providing inertial response and frequency control.

\section{Slack Bus as a Set of Differential Equations} 
\label{sec:theory}

Equation \eqref{eq:1} is commonly interpreted as an algebraic (static) constraint.  In this work, on the other hand, we interpret \eqref{eq:1} as the equilibrium of a perfect tracking controller.  In its simplest form, the dynamic form of \eqref{eq:1} can be assumed to be a pure integrator:
\begin{equation}
  \label{eq:int}
  \hat{\sigma}' = \theta^o - \theta_i \, .
\end{equation}
Equations \eqref{eq:1} and \eqref{eq:int} are equivalent as, in steady state, $\dot{\sigma} = 0$.  However, interpreting the slack as a differential equation allows further generalizing the concept of slack bus, as follows.


Equation \eqref{eq:int} can be a non-perfect tracking controller.  For example, a first-order droop control can be written as:
\begin{equation}
  \label{eq:slack}
  T \hat{\sigma}' = K (\theta^o - \theta_i) - H \hat{\sigma} \, ,
\end{equation}
For the power flow analysis, \eqref{eq:1}, \eqref{eq:int} and \eqref{eq:slack} are equivalent, except for the fact that, if $H \ne 0$, \eqref{eq:slack} leads to $\theta_i \ne \theta^o$ in steady state.  This difference is immaterial as the solution of any ac circuit represented in phasor domain is unique up to a phase angle reference.


Moreover, the dynamics of the slack variable do not need to be a first order nor a linear differential equation.  As long as they depend on $\theta_i$, one can thus assume any set of differential equations for the slack variable(s):
\begin{equation}
  \label{eq:slackvec}
  \bfb T \bfg \sigma' = \bfg f(\bfg \sigma, \theta_i) \, , 
\end{equation}
where $\bfg \sigma \in \mathbb{R}^m$, $\bfg f: \mathbb{R}^{m+1} \mapsto \mathbb{R}^{m}$ and $\bfb T \in \mathbb{R}^m \times \mathbb{R}^m$.  Some rows of $\bfb T$ can be null, thus leading to a set of differential-algebraic equations.

Observing \eqref{eq:2}, we note that the term $k_{g,h} \sigma$ has the units of a power.  In general, this power can be composed of two terms; one term that is not null in steady-state, say $p_{h,s}$; and one transient term that vanishes in steady-state conditions, say $p_{h,t}$, as follows:
\begin{equation}
  \label{eq:L}
  p_h = p_{h,s}(\bfg \sigma) + p_{h,t}(\bfg \sigma, \bfg \sigma') \, .
\end{equation}
where we have assumed that $p_{h,s}$ depends only on the states $\bfg \sigma$ and $p_{h,t}$ on both the state and their first time derivative.  The latter dependency is what makes $p_{h,t}$ vanish in steady-state.  For example, matching \eqref{eq:2} with \eqref{eq:L}, one has $p_{h,s} \equiv p^o_h + k_{g,h} \hat{\sigma}$ and $p_{h,t} \equiv 0$.  Equation \eqref{eq:L} indicates that a device has slack bus capability only if it has a power source ($p_{s,h}$), as expected.  On the other hand, the transient power component, which is due to some form of stored energy ($p_{t,h}$), vanishes in steady state.  The case study further elaborates on this point and discusses the role of the transient component in the relevant cases of synchronous machines, as well as of GFL and GFM converters.


Finally, we note that there can be multiple (local) constraints on the bus voltage phase angles.  One can in fact assume that each generator has its own local set of slack differential equations plus active power injection.


Combining the remarks above, we propose the following definition of device providing slack bus capability.

\vspace{1mm}

\noindent \fbox{
  \begin{minipage}{0.95\linewidth}
    \noindent \textbf{Definition:}  A set of differential-algebraic equations in the form:
\begin{align}
      \label{eq:slackdis}
      \bfb T_h \bfg \sigma'_h &= \bfg f_h(\bfg \sigma_h, \theta_h) \, , \\
      \label{eq:gen}
      p_{h} &= p_{s,h}(\bfg \sigma_h) + p_{t,h}(\bfg \sigma_h, \bfg \sigma_h') 
\end{align}
has \textit{slack bus capability} if, at equilibrium, it satisfies the condition:
\begin{equation}
  \label{eq:nondeg1}
  \lim_{t \rightarrow \infty} \sigma_{k,h} \rightarrow \hat{\sigma} \, , \quad \exists \, k \in (1, \dots, m_h), \; \; \forall \, h \in (1, \dots, n) \, .
\end{equation}
\end{minipage}
}

\vspace{1mm}

In \eqref{eq:slackdis}, $\bfg \sigma_h \in \mathbb{R}^{m_h}$, $\bfg f_h: \mathbb{R}^{m_h+1} \mapsto \mathbb{R}^{m_h}$ and $\bfb T_h \in \mathbb{R}^{m_h} \times \mathbb{R}^{m_h}$; $p_{h}$ is the active power injected by the device into the grid;  and $\theta_h$ is the bus voltage phase angle at a certain bus of the grid, generally, but not necessarily, the angle of the bus where the device is connected.  In \eqref{eq:nondeg1}, $\hat{\sigma} \in \mathbb{R}$ is a common value for all devices providing slack bus capability.

Condition \eqref{eq:nondeg1} means that, in steady state, local constraints \eqref{eq:slackdis} converge to the distributed slack bus model as in \eqref{eq:2}.  Note, however, that the condition \eqref{eq:1} is not required.  Each device providing slack bus capability will determine its local $\theta_h$ that satisfies the steady-state condition $\bfg f_h(\bfg \sigma_h, \theta_h) = \bfg 0$.

Equations \eqref{eq:slackdis}-\eqref{eq:gen} represent a distributed, multi-variable, dynamic local slack bus model: \textit{distributed} as there the slack variables appear in the power injection of each generator; \textit{multi-variable} as $\bfg \sigma_h$ is a vector or order greater than one; \textit{dynamic} as $\bfg \sigma'_h \ne \bfg 0$ in general; and \textit{local} as each generator determines the local bus voltage phase angle $\theta_h$.  For comparison, model \eqref{eq:2} and \eqref{eq:slack} is distributed, single-variable, dynamic, network-wide slack bus.  And model \eqref{eq:1} and \eqref{eq:2} with all $k_{g,h} = 0$ except one is a centralized, single-variable, static, network-wide slack bus.  Any other ``combination'' is equally valid as long as \eqref{eq:nondeg1} is satisfied.  This is illustrated below through a variety of examples.

\vspace{-2mm}
\section{Examples}

\subsection{Synchronous Machines}

In light of the definition above, the classical swing equation of the synchronous machine can be interpreted as a distributed, multi-variable dynamic slack bus of the kind \eqref{eq:slackdis}-\eqref{eq:gen}.  In fact, the swing equations of the machine are in the form of \eqref{eq:slackdis}, as follows:
\begin{equation}
  \label{eq:swing}
  \begin{aligned}
    \delta'_h &= \Omega_b \, (\omega_h - \omega_s) \, , \\
    M_h \, \omega'_h &= \tau_{m, h} - \tau_{e, h}^{\max} \sin(\delta_h - \theta_h) - D_h \, (\omega_h -\omega_n) \, ,
  \end{aligned}
\end{equation}
where $\tau_{e, h}^{\max}$
is the maximum electrical torque;  $\tau_{m, h}$ is the mechanical torque; $\omega_n$ is the nominal synchronous speed with, typically, $\omega_n = 1$ pu;  $\delta_h$ and $\omega_h$ are the angle and the rotor speed, respectively; $\omega_s$ is the angular speed of a common reference angular speed (typically, the angular speed of the center of inertia); $\Omega_b$ is the angular frequency base in rad/s; and parameters have the usual meaning, namely,  $M_h$ is the mechanical starting time; and $D_h$ is the damping~\cite{Milano:2010}.   Hence, in this model, $\bfg \sigma = [\delta_h, \omega_h]^T$.  The power injection of the machine into the grid is in the form of \eqref{eq:gen}, where  the two components of the power are:
\begin{equation}
  \label{eq:psg1}
  p_{s, h} = \tau_{m,h} \, \omega_h - D_h \, (\omega_h - \omega_n)^2 \, , 
\end{equation}
where $\tau_{m,h}$ is assumed to be constant for the classical machine model.  If $\omega_h \approx \omega_s \approx 1$ pu, then one obtains $p_{m,h} \approx \rm const.$, hence:
\begin{equation}
  \label{eq:psg}
  p_{s, h} = p^o_{m,h} - D_h \, (\omega_h - \omega_n)^2  \, ,
\end{equation}
and
\begin{equation}
  \label{eq:ptg}
  p_{t, h} = -M_h \omega'_h \omega_h = -\frac{d}{dt} \left ( \frac{1}{2} M_h \omega_h^2 \right) \, .
\end{equation}
The last expression shows that the transient power is due to the kinetic energy stored in the machine inertia.  Moreover, if $\omega_h \approx 1$,  $p_{t, h} \approx M_h \omega'_h$, which is the term that appears in the swing equations of the machine, namely \eqref{eq:swing}.
More detailed models of the synchronous machine includes more states, such as transient and sub-transient rotor fluxes but, ultimately, all models can be written in the form of \eqref{eq:slackdis}, that is, depends on states and $\theta_h$.

As the system converges to an equilibrium, if $D_h\ne 0$ at least for some machines, then
\begin{equation}
  \lim_{t\rightarrow \infty} \omega_h \rightarrow \omega_{s} \, ,
\end{equation}
This satisfies the condition \eqref{eq:nondeg1}.  It is relevant to observe that, if the damping is $D_h = 0$ for all machines and there is no frequency regulation, after a contingency, the system either diverges or enters into a stationary periodic motion, which does not satisfy \eqref{eq:nondeg1}.  Yet, the average value of the rotor speeds of the machines still satisfies \eqref{eq:nondeg1}.  We can thus define a \textit{weak slack bus} condition as follows:
\begin{equation}
  \label{eq:nondeg2}
  \lim_{t \rightarrow \infty} \langle \sigma_{k,h} \rangle \rightarrow \hat{\sigma} \, , \quad \exists \, k \in (1, \dots, m_h), \quad \forall \, h \in (1, \dots, n) \, ,
\end{equation}
where $\langle \cdot \rangle$ represents the average over the period of the motion.

\vspace{-2mm}
\subsection{Primary Frequency Control}

In this example, we consider synchronous machines with primary frequency control.  In its simplest form, the model of a turbine governor can be written as:
\begin{equation}
  \label{eq:primary}
  \begin{aligned}
    T_h \, \tau'_{m, h} &= \tau^o_{m, h} - \frac{1}{R_h} (\omega_h - \omega_n) - \tau_{m, h} \, ,
  \end{aligned}
\end{equation}
where $R_h$ is the droop coefficient,  $T_h$ is the time constant of the primary frequency control, and $p^o_{m, h}$ is the turbine power set point.  Thus, \eqref{eq:psg} can be rewritten as:
\begin{equation}
  \label{eq:psgov}
  p_{s,h} = p^o_{m, h} - (D_h + R^{-1}_h) \, (\omega_h - \omega_n)^2 \, ,
\end{equation}
where $p^o_{m, h} = \tau^o_{m, h} \, \omega_h \approx \tau^o_{m, h} \, $.
The transient term is:
\begin{equation}
  \label{eq:ptgov}
  \begin{aligned}
    p_{t,h} &= - M_h \, \omega'_h \, \omega_h - T_h \, \tau'_{m,h} \, \omega_h \\
    &\approx - M_h \, \omega'_h - T_h \, \tau'_{m,h} \, ,
  \end{aligned}
\end{equation}
where, for slow dynamic of the governor, the second term is small with respect to the inertial response of the synchronous machine.
The following remarks are relevant.

\subsubsection*{Remark 1}

The effect of the droop frequency control is, ultimately, to modify the damping of the machine.  Thus, in steady state, the expression of $p_{s,h}$ in \eqref{eq:psgov} satisfies the slack bus condition \eqref{eq:nondeg1}.

\subsubsection*{Remark 2}

If one turbine governor is integral, for example:
\begin{equation}
  \label{eq:primary2}
  \begin{aligned}
    T_h \, \tau'_{m, h} &= - \frac{1}{R_h} (\omega_h - \omega_n) \, ,
  \end{aligned}
\end{equation}
then, in steady state, $(\omega_h - \omega_n)^2 = 0$, which leads to $p_{s,h} = p^o_{m, h}$ for all machines except for the one with integral frequency control.  For this machine, $p_{s,h}$ is whatever active power satisfies the power balance at the point of connection of the machine with the grid.  This case is equivalent to the single slack bus model \eqref{eq:1}.  Moreover, only one turbine governor can be integral for an interconnected grid, otherwise, the active power injections of the generators connected to the integral governors are undetermined.  This conclusion is generally obtained in the context of primary frequency control (see, e.g., \cite{Milano:2010}) not as a particular model of dynamic slack bus.

\vspace{-2mm}
\subsection{Automatic Generation Control}

It is possible to show that a perfect-tracking automatic generation control (AGC) is equivalent in steady state to a distributed slack bus model.  This concept has been discussed in the literature (see, e.g., \cite{Dhople20}), however, the derivation of this result in the framework of a generalized dynamic slack bus model is a contribution of this paper.  In particular, we show, based on proposed definition, that the ensemble of an AGC coupled with turbine governors and synchronous machines constitutes a distributed, multi-variable, dynamic, network-wide slack bus model.

The simplest AGC model is an integral controller:
\begin{equation}
  \label{eq:agc}
  \xi' = K_o (\omega_n - \omega_s) \, ,
\end{equation}
where $\xi$ and $K_o$ are the state variable and the gain, respectively, of the AGC integrator.  The AGC output signal $\xi$ is shared with the turbine governors of the synchronous machines and adjusts their power set point.  Equation \eqref{eq:primary} is thus rewritten as:
\begin{equation}
  \label{eq:tgagc}
  T_h \, \tau'_{m, h} = \tau^o_{m, h} + r_h \, \xi - \frac{1}{R_h} (\omega_h - \omega_n) - \tau_{m, h} \, ,  
\end{equation}
where, in conventional implementations, $r_h = R_h \, (\sum^n_{h=1} R_h)^{-1}$.  As \eqref{eq:agc} is perfect tracking, in steady state, the model composed of \eqref{eq:swing}, \eqref{eq:agc} and \eqref{eq:tgagc} leads to the following active power injection for the $h$-th synchronous machine:
\begin{equation}
  \label{eq:pgagc}
  p_{h} = p^o_{m, h} + \tilde{r}_h \, \xi \, ,
\end{equation}
where $\tilde{r}_h = r_h \omega_n$ and which has the same form as \eqref{eq:2} --- and thus satisfies \eqref{eq:nondeg1} --- with $\hat{\sigma} \equiv \xi$.

\vspace{-2mm}
\subsection{Passive Load}

This example considers a passive load modeled as a series RLC circuit.  Using the $\rm dq$-axis reference frame, the load equations at bus $h$ are:
\begin{equation}
  \label{eq:line}
  \begin{aligned}
    \ell_{h} \, \bar{\imath}'_{\ell,h} &= \bar{v}_h - \bar{v}_{c,h} - (r_{h} + j \Omega_b \ell_{h}) \,  \bar{\imath}_{\ell, h} \, , \\
    c_{h} \, \bar{v}'_{c,h} &= \bar{\imath}_{\ell, h} - j \Omega_b c_{h} \, \bar{v}_{c,h}  \, ,
  \end{aligned}
\end{equation}
where $\bar{v}_h = v^{\rm d}_h + j v^{\rm q}_h$, $\bar{v}_{c,h} = v^{\rm d}_{c,h} + j v^{\rm q}_{c,h}$, $\bar{\imath}_{\ell, h} = i^{\rm d}_{\ell, h} + j i^{\rm q}_{\ell, h}$ are the Park vectors of the voltage at the grid bus, the voltage on the capacitor and the load current, respectively;  $r_{hk}$, $j \Omega_b \ell_{hk}$ and $j \Omega_b c_{h}$ are the resistance, inductive reactance and capacitive susceptance, respectively, of the load.  Then, the source and transient powers are:
\begin{equation}
  \label{eq:psptline}
  \begin{aligned}
    p_{s,h} &= 0 \, , \\
    p_{t,h} &= \ell_{hk} i'_{hk} i_{hk} + c_{h} v'_{h} v_{h} \, ,
  \end{aligned}
\end{equation}
where $p_{s,h}$ is zero as the load is a passive device and $p_{t,h}$ accounts for the energy stored in the capacitive and inductive elements.  The load, as expected and as any other passive element of the grid, including transmission lines, cannot provide any steady-state slack bus capability, and its transient support runs out quickly as it is due to the energy stored in its inductive and capacitive elements.

\vspace{-2mm}

\subsection{Grid-Following Converters}

Grid-following converters (GFLs) are synchronized to the ac grid through a phase-locked loop (PLL), and include a dc circuit, an ac filter as well as inner-loop current controllers and outer-loop voltage controllers.  A detailed description of typical dynamic models of GFL converters for system stability studies can be found, for example, in \cite{Milano2019, Dhople23}.
The active power injected by the GFL into the grid is:
\begin{equation}
  \label{eq:pdq}
  \begin{aligned}
    p_{h} &= v^{\rm d}_{h} \, i^{\rm d}_{h} + v^{\rm q}_{h} \, i^{\rm q}_{h} \, ,
  \end{aligned}
\end{equation}
where the current $\rm dq$-axis components, namely  $(i^{\rm d}_{h}, i^{\rm q}_{h})$, are controlled in order to impose a reference dc and ac voltage; and the converter voltage $dq$-axis components $(v^{\rm d}_{h}, v^{\rm q}_{h})$ are linked  to the grid voltage $(v_h, \theta_h)$, as follows:
\begin{equation}
  \label{eq:vgfl}
  \begin{aligned}
    v^{\rm d}_{h} &= v_h \cos(\theta_h - \hat{\theta}_h) \, , \\
    v^{\rm q}_{h} &= v_h \sin(\theta_h - \hat{\theta}_h) \, , 
  \end{aligned}
\end{equation}
where $\hat{\theta}_h$ is the angle estimated through the PLL, a simple implementation of which is:
\begin{equation}
  \label{eq:pll}
  \begin{aligned}
    \zeta'_h &= \theta_h - \hat{\theta}_h \, , \\
    -k^{\rm p}_h \, \zeta'_h &= k_h^{\rm i} \, \zeta_h - \Delta \hat{\omega}_h \, , \\
    \hat{\theta}'_h &= \Omega_b \, \Delta \hat{\omega}_h \, , 
  \end{aligned}
\end{equation}
where $\zeta_h$, $k^{\rm p}_h$ and $k^{\rm i}_h$ are the integrator state, proportional gain, and integral gain, respectively, of the PLL loop filter; and $\Delta \hat{\omega}_h$ is the estimated frequency deviation with respect to the synchronous speed at the point of connection of the converter at the grid. 

Then, the power injected into the grid can be decomposed into the following source and transient active power expressions:
\begin{equation}
  \label{eq:p_gfl}
  \begin{aligned}
    p_{s, h} &= p_{\rm dc} - g_{\rm dc} \, v_{\rm dc}^2 - r_{\rm f} \, i_h^2 \, , \\
    p_{t, h} &= c_{\rm dc} \, v'_{\rm dc} v_{\rm dc} + \ell_{\rm f} \, i_h' i_h + c_{\rm f} \, v'_h v_h \, ,
  \end{aligned}
\end{equation}
where $r_{\rm f}$, $\ell_{\rm f}$, $c_{\rm f}$ are the resistance, inductance and capacitance (expressed in pu) of the ac filter of the converter; $g_{\rm dc}$ is an equivalent conductance that accounts for the losses of the converter; $c_{\rm dc}$ is the condenser utilized to filter the ripple on the converter dc voltage; $v_h$ and $i_h$ are the ac voltage and current, respectively;  $v_{\rm dc}$ and $i_{\rm dc}$ are the dc voltage and current, respectively; and $p_{\rm dc} = v_{\rm dc} i_{\rm dc}$ is the dc power injected into the converter.  From \eqref{eq:p_gfl}, one obtains that the GFL can  provide slack bus capability in steady-state only if the dc side includes a controlled energy source.  For example, if the dc current of the energy source is regulated through a droop frequency controller, one has:
\begin{equation}
  T_{\rm dc} i'_{\rm dc} = i^o_{\rm dc} -\frac{1}{R_{\rm dc}} (\omega_h - \omega_n) - i_{\rm dc} \, ,
\end{equation}
which leads to rewrite \eqref{eq:p_gfl} as:
\begin{equation}
  \label{eq:pgfl2}
  \begin{aligned}
    p_{s, h} &= v_{\rm dc} i^o_{\rm dc} -\frac{v_{\rm dc}}{R_{\rm dc}} (\omega_h - \omega_n) - g_{\rm dc} \, v_{\rm dc}^2 - r_{\rm f} \, i_h^2 \, , \\
               p_{t, h} &= T_{\rm dc} i'_{\rm dc}i_{\rm dc} + c_{\rm dc} \, v'_{\rm dc} v_{\rm dc} + \ell_{\rm f} \, i_h' i_h + c_{\rm f} \, v'_h v_h \, ,
  \end{aligned}
\end{equation}
where $i^o_{\rm dc}$ is the initial steady-state value of the dc current; and $T_{\rm dc}$ and $R_{\rm dc}$ are the time constant and the droop coefficient of the primary frequency controller of the GFL.
In steady state, the first equation of \eqref{eq:pgfl2} resembles the distributed slack model \eqref{eq:2}.
Moreover, similarly to the case of the passive load, the transient power $p_{t,h}$ is due to the energy stored in the dc and ac filters, and it decays in the order of milliseconds in case of a disturbance.

\vspace{-2mm}

\subsection{Grid-Forming Converters}

Grid-forming converters (GFMs) can be formulated as a set of equations similar to \eqref{eq:swing}-\eqref{eq:ptg} \cite{rosso2021grid}.  The main difference with GFLs, is how the converter voltage is linked to the grid voltage, as follows:
\begin{equation}
  \label{eq:vgfm}
  \begin{aligned}
    v^{\rm d}_{h} &= v_h \cos(\alpha_h - \hat{\theta}_h) \, , \\
    v^{\rm q}_{h} &= v_h \sin(\alpha_h - \hat{\theta}_h) \, , 
  \end{aligned}
\end{equation}
where the variable $\alpha_h$ is utilized in the active power control of the GFM:
\begin{equation}
  D_{\alpha} \alpha'_h = p_h^{\rm ref} - p_{h} - H_\alpha \, \alpha_h \, ,  
\end{equation}
where $p_h^{\rm ref}$ is the reference active power and $D_{\alpha}$ and $H_\alpha$ are the regulator time constant and integral deviation, respectively.
from where one can deduce that the source and transient powers of the GFM (omitting for simplicity the terms due to dc and ac filters and losses which are similar to the ones shown for the GFL) are:
\begin{equation}
  \begin{aligned}
    p_{s,h} &= p_{\rm dc} - p_{\rm losses}  - H_\alpha \, \alpha_h \, ,  \\
    p_{t,h} &= D_{\alpha} \, \alpha_h' \, 
  \end{aligned}
\end{equation}
where $v_{\rm dc} i_{\rm dc} - p_{\rm losses} = p_h^{\rm ref}$ in steady state, thus resembling the case of a distributed slack \eqref{eq:2}.  In some configurations, the dynamics of the GFM are implemented in such a way that the converter behaves as a virtual synchronous machine \cite{arco:2017}, leading to:
\begin{equation}
  \begin{aligned}
    \alpha'_h
    &= \tilde{\omega}_{h} \, , \\
      M_{\alpha} \tilde{\omega}_{h} '
    &= p_h^{\rm ref} - p_{h} - D_{\alpha} \, \tilde{\omega}_h - H_{\alpha} \, \alpha_h \, ,
  \end{aligned}
\end{equation}
where $M_{\alpha}$ is a virtual inertia constant, and
\begin{equation}
  \begin{aligned}
    p_{s,h} &= p_{\rm dc} - p_{\rm losses} - H_{\alpha} \, \alpha_h - D_{\alpha} \, \tilde{\omega}_{h} \, ,  \\
    p_{t,h} &= M_{\alpha} \, \tilde{\omega}'_{h} \, ,
  \end{aligned}
\end{equation}
which closely resemble \eqref{eq:psg} and \eqref{eq:ptg} of synchronous machines.

\vspace{-2mm}

\section{Case Study}
\label{sec:CaseStudy}

This section illustrates the proposed framework of dynamic slack bus through a numerical case study.  With this aim, we consider the WSCC 9-bus test system and consider the following scenarios: (i) base case system with 3 synchronous machines, AVRs and turbine governors; (ii) and (iii) scenarios where the synchronous machines are substituted with three GFMs converters; and (iv) a scenario where the synchronous machines are substituted with three GFLs converters.  For scenarios (ii) and (iii), two GFM models are considered, namely, REGFM\_A1 \cite{REGFM_A1} and REGFM\_B1 \cite{REGFM_B1}, respectively.  All scenarios include primary frequency and voltage control.  For all scenarios, the contingency is a loss of 20\% of the load consumption at bus 5 at $t=1$ s.  Simulations are carried out with the software tool Dome \cite{Dome}.

\begin{figure}[ht!]
  \begin{center}
    \resizebox{0.95\linewidth}{!}{\includegraphics[scale=1.0]{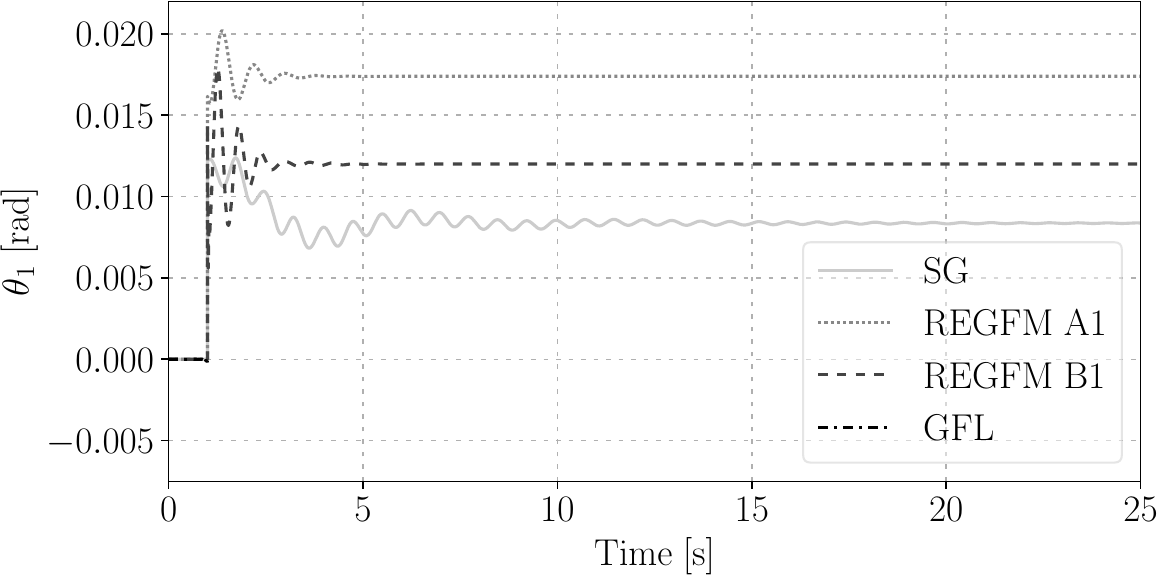}}
    \caption{Trajectories of the voltage phase angle at bus 1 ($\theta_1$) of the WSCC 9-bus system for different generation scenarios.}
    \label{fig:sim}
  \end{center}
  \vspace{-3mm}
\end{figure}

Figure \ref{fig:sim} shows the trajectory of the voltage phase angle at bus 1 for the four scenarios.  The phase angles of the voltages at the other buses show a similar transient behavior.  The initial operating condition is the same and is stable for all scenarios.  The conventional synchronous machines, as well as the two models of GFMs have enough stored energy to compensate the power balance during the transient.  On the other hand, in scenario (iv), the GFLs do not have enough stored energy to cope with the transient conditions and the simulation stops instants after $t=1$ s.  This result was to be expected, as GFLs do not provide an inertial response.

\vspace{-2mm}

\section{Conclusions}
\label{sec:Conclusion}

The letter shows that the well-known distributed slack bus model can be interpreted as the steady-state version of a set of differential-algebraic equations.  The dynamic version includes a transient component which has a role in the \textit{weak} version of the proposed definition.  A variety of examples applies the proposed framework to a variety of devices, including passive loads and electronic power converters with both grid-following and grid-forming controls.  Future work will elaborate on the proposed framework and study whether it is possible to set up devices with dynamic slack bus capability where the slack variables do not converge, in steady-state, to the system frequency.  The author also aims at utilizing the proposed mathematical framework to define new mechanisms for converter-based resources to provide power balance support to the grid.



\end{document}